\documentclass[twocolumn]{jpsj3}
\usepackage{txfonts}
\usepackage{multicol}
\usepackage{times}
\usepackage{color}
\usepackage{bm}
\usepackage{subfigure}

\title{
Electric Octupole Order in Bilayer Ruthenate Sr$_3$Ru$_2$O$_7$ 
}

\author{Takanori Hitomi$^{1}$ and 
Youichi Yanase$^{1,2}$\thanks{E-mail: yanase@phys.sc.niigata-u.ac.jp}
}
\inst{
$^{1}$Graduate School of Science and Technology, Niigata University, Niigata 950-2181, Japan
\\
$^{2}$Department of Physics, Niigata University, Niigata 950-2181, Japan
}

\recdate{July 7, 2014}

\abst{
Although the odd-parity multipole order barely occurs in crystals with high symmetry, it can be formed 
in locally noncentrosymmetric crystals. 
We illustrate the odd-parity electric octupole order generated in a bilayer structure. 
When the local electric quadrupole is alternatively stacked between layers, it is regarded as an electric octupole order 
from the viewpoint of symmetry. We show that the $p_y s_x + p_x s_y$ spin nematic order is induced by the spin-orbit coupling 
in the electric octupole state, and it results from the spontaneous inversion symmetry breaking leading to the $D_{2d}$ point group symmetry. 
We investigate the possible realization of the electric octupole order in the bilayer ruthenate Sr$_3$Ru$_2$O$_7$, 
assuming a local electric quadrupole arising from the Pomeranchuk instability and/or the orbital order. 
Effects of the lattice distortion and magnetic field are also clarified, and the nature of the ``electronic nematic state'' 
of Sr$_3$Ru$_2$O$_7$ is discussed. 
It is proposed that the asymmetric band structure is a signature of the electric octupole order in Sr$_3$Ru$_2$O$_7$. 
The odd-parity multipole order in other strongly correlated electron systems is discussed. 
}

\begin{document}
\maketitle

\renewcommand{\k}{{\bm k}}
\renewcommand{\r}{{\bm r}}
\newcommand{\dd}{{\bm d}}
\newcommand{\kk}{{\bm k'}}
\newcommand{\kkk}{{\bm k''}}
\newcommand{\q}{{\bm q}}
\newcommand{\Q}{{\bm Q}}
\newcommand{\e}{\varepsilon}
\newcommand{\ee}{e}
\newcommand{\s}{{\mit{\it \Sigma}}}
\newcommand{\J}{\mbox{\boldmath$J$}}
\newcommand{\vv}{\mbox{\boldmath$v$}}
\newcommand{\Jh}{J}
\newcommand{\LL}{\mbox{\boldmath$L$}}
\renewcommand{\SS}{\mbox{\boldmath$S$}}
\newcommand{\MM}{\mbox{\boldmath$M$}}
\newcommand{\g}{\mbox{\boldmath$g$}}
\newcommand{\HH}{\mbox{\boldmath$H$}}
\newcommand{\hh}{\mbox{\boldmath$h$}}
\newcommand{\Tc}{$T_{\rm c}$ }
\newcommand{\Tcf}{$T_{\rm c}$}
\newcommand{\Hc}{$H_{\rm c2}$ }
\newcommand{\Hcf}{$H_{\rm c2}$}
\newcommand{\etal}{{\it et al.}: }
\newcommand{\SRO}{Sr$_2$RuO$_4$ }
\newcommand{\SROf}{Sr$_2$RuO$_4$}
\newcommand{\px}{p_{x}}
\newcommand{\py}{p_{y}}

\section{Introduction}

 Exotic quantum phases induced by the spin-orbit coupling have been attracting recent interest 
in various fields of condensed matter physics, owing to their fascinating phenomena  
such as noncentrosymmetric superconductivity~\cite{NCSC}, 
chiral magnetism~\cite{Chiral_Magnetism_1,Chiral_Magnetism_2}, multipole order~\cite{Multipole}, 
multiferroics~\cite{Katsura}, spintronics~\cite{Spin_Hall_1, Spin_Hall_2}, 
and topological quantum phases~\cite{RevModPhys.82.3045,RevModPhys.83.1057,Tanaka_review}. 
Intriguing properties appear in the electronic structure, transport, optics, magnetic excitation, and so forth, 
as a result of the spin-orbit coupling and symmetry breaking. 

 In this paper, we illustrate a novel multipole order accompanied by the spontaneous inversion symmetry breaking.  
Although various high-rank multipole orders have been proposed for d- and f-electron systems~\cite{Multipole}, 
previous studies have focused on the even-parity multipole, such as the electric quadruple and the magnetic octupole. 
We can define the counterpart, namely, the odd-parity multipole. However, it has not been considered 
because it is barely polarized from the electronic degree of freedom. 
For instance, the ferroelectric excitonic state~\cite{Excitonic_phase} is the lowest-rank odd-parity multipole state, 
but we have not found the excitonic order in materials so far.

 On the other hand, the odd-parity multipole can be formed in locally noncentrosymmetric crystals 
without relying on the exotic electron correlation effect. 
We outline the mechanism as follows.
A unit cell of the crystals contains several atoms, and the electrons have a sublattice degree of freedom. 
The odd-parity multipole is defined in the unit cell, and therefore induced by the site-dependent even-parity multipole. 
For instance, the magnetic quadrupole order is induced by the ``antiferromagnetic'' order in 
zigzag chains.~\cite{Magnetic_Quadrupole} Similarly, the toroidal order is induced in the honeycomb lattice~\cite{Hayami_1}. 
In these states, the $p$-wave charge nematic order is induced by the spin-orbit coupling~\cite{Magnetic_Quadrupole}. 
In this paper, we study its natural extension, that is, the electric octupole (EO) order formed by the 
``antiferro" alignment of the local electric quadrupole (EQ) in the unit cell. 
It is shown that the $p$-wave spin nematic order is induced by the spin-orbit coupling.

 For demonstration, we introduce a model for the EO order, on the basis of which its possible realization 
in the bilayer ruthenate Sr$_3$Ru$_2$O$_7$~\cite{Ikeda, Sr3Ru2O7_nematic_1} is discussed. 
The hidden-ordered state appears in Sr$_3$Ru$_2$O$_7$ at high magnetic fields near the quantum critical point.
As evidenced by several experimental results,~\cite{Sr3Ru2O7_nematic_2, Sr3Ru2O7_nematic_3} 
it is considered to be an ``electronic nematic state''. 
However, its nature and origin are still under debate~\cite{Sr3Ru2O7_review}. 
Two possible scenarios have been theoretically proposed.  
One is the $d$-wave Pomeranchuk instability (dPI) of the $d_{xy}$-orbital~\cite{Kee-Kim,Yamase-Katanin,Yamase,Adachi}, 
and the other is the orbital order caused by the degenerate ($d_{yz}$, $d_{zx}$)-orbitals~\cite{Raghu,Lee-Wu,Ono,Tsuchiizu}. 
When the single-layer model is adopted as in early studies, both ordered states are classified into 
the same electric quadrupole ($O_{x^2-y^2}$) state. Indeed, the two order parameters of dPI and orbital order are admixed. 
On the other hand, we can consider the two multipole states distinguished by the symmetry 
when we take into account the bilayer structure in the unit cell~\cite{Puetter_bilayer,Yamase_bilayer}. 
One is the ferro stacking of quadrupoles in the bilayer, and the other is the antiferro stacking. 
When we regard the unit cell as a multimer, the former is classified into the EQ order, 
and the latter is the  EO order that we focus on.

We investigate the electronic structure of Sr$_3$Ru$_2$O$_7$ in the EQ and EO states. 
In particular, we show that the $p_y s_x + p_x s_y$ spin nematic order is induced by the spin-orbit coupling in the EO state. 
The induced spin nematic order can be regarded as the emergence of the antisymmetric spin-orbit coupling 
through the spontaneous inversion symmetry breaking due to the odd-parity multipole order. 
We clarify the spin texture symmetry in the momentum space on the basis of the point group symmetry,  
and numerically examine the difference between the dPI-dominated EO state and the orbital-order-dominated EO state. 
It is also pointed out that the local noncentrosymmetricity of the bilayer structure plays essential roles. 
Since the odd-parity hybridization terms arising from the local violation of crystal mirror symmetry have been 
neglected in previous theories~\cite{Puetter_bilayer,Yamase_bilayer}, the exotic electronic structure of 
the EO state has been overlooked.

In Sect.~2, we construct the $3 \times 2 \times 2 =12$ component tight-binding model, 
taking into account the three $t_{\rm 2g}$-orbitals, two layers, and two spin states. 
The EQ and EO states are investigated by adopting the mean field terms 
representing the ferro and antiferro dPI and/or orbital order, respectively. 
We show the signatures of the dPI and orbital order and propose an experimental test for these scenarios. 
In Sect.~3, the spin texture in the  EO state is elucidated. 
The spin texture symmetry and induced spin nematic order are clarified on the basis of 
the $D_{2d}$ point group symmetry of the EO state. 
In Sect.~4, we study the effect of the lattice distortion due to the rotation of RuO$_6$ octahedra. 
In Sect.~5, 
we show that the asymmetric band structure appears 
in the magnetic field and its field angle dependence would be a signature of the  EO order. 
In Sect.~6 we summarize the results and discuss other odd-parity multipole orders 
in locally noncentrosymmetric systems.

\section{Electronic Structures in Electric Quadrupole and Electric Octupole States}

\subsection{Mean field model for Sr$_3$Ru$_2$O$_7$ in the nematic state}

First, we introduce a bilayer three-orbital tight-binding model for Sr$_3$Ru$_2$O$_7$. 
The model is obtained by a straightforward extension of the single-layer model for the surface state 
of Sr$_{2}$RuO$_{4}$~\cite{a} to the bilayer Sr$_3$Ru$_2$O$_7$. 
Although the nematic order occurs in the magnetic 
field,~\cite{Sr3Ru2O7_nematic_2, Sr3Ru2O7_nematic_3}  here, we neglect the magnetic field to clarify 
the electronic structure in the EO state. We will show the intriguing effect of the magnetic field in Sect.~5.
The model is given by
\begin{align}
\label{H_0}
H_{\rm{0}}&=H_{\rm{kin}} + H_{\rm{hyb}} + H_{\rm{CEF}} + H_{\rm{odd}} + H_{\rm{LS}} + H_{\perp} + H_{\rm{lp}}, \\
H_{\rm{kin}}&=\sum_{\bm{k}} \sum_{m=1,2,3} \sum_{s=\uparrow,\downarrow} \sum_{l=A,B} \varepsilon_m (\bm{k}) c_{\bm{k}msl}^{\dagger} c_{\bm{k}msl},\\
H_{\rm{hyb}}&=\sum_{\bm{k},s,l} [ V(\bm{k}) c_{\bm{k}1sl}^{\dagger} c_{\bm{k}2sl} + \rm{h.c.} ],\\
H_{\rm{CEF}}&=\Delta \sum_{\bm{k},s,l} c_{\bm{k}3sl}^{\dagger} c_{\bm{k}3sl},\\
H_{\rm{odd}}&=\sum_{\bm{k},s,l} [ V_{x, \, l}(\bm{k}) c_{\bm{k}1sl}^{\dagger} c_{\bm{k}3sl} + V_{y, \, l}(\bm{k}) c_{\bm{k}2sl}^{\dagger} c_{\bm{k}3sl} + \rm{h.c.}],\\
H_{\rm{LS}}&=\lambda \sum_{i} \sum_{l} \bm{L}_{il} \cdot \bm{S}_{il},\\
H_{\perp}&=\sum_{\bm{k},m,s} t_{\perp, \, m} [c^{\dagger}_{\bm{k}msA} c_{\bm{k}msB} + \rm{h.c.}], \\
H_{\rm{lp}}&=g_{\rm{lp}} \sum_{\bm{k},m,s,l}  [c^{\dagger}_{\bm{k}msl} c_{\bm{k}+\bm{Q}msl} + \rm{h.c.}],
\end{align}
where $c_{\bm{k}msl}$ ($c^{\dagger}_{\bm{k}msl}$) is the annihilation (creation) operator of an electron with the orbital 
$m=1,2,3$ and spin $s=\uparrow,\downarrow$ on the layer $l=A,B$. We denote $(d_{yz}, d_{zx}, d_{xy})$-orbitals of Ru ions 
using the index $m=(1,2,3)$, respectively. 
The first term $H_{\rm{kin}}$ is the intra-orbital kinetic energy term. The second term $H_{\rm{hyb}}$ describes the 
hybridization between the $d_{yz}$- and $d_{zx}$-orbitals through the hopping between the next-nearest-neighbor Ru sites. 
The third term $H_{\rm{CEF}}$ introduces the crystal electric field with tetragonal symmetry. 
Since the local mirror symmetry is broken on the RuO$_{2}$ planes, the $d_{xy}$-orbital and $(d_{yz} , d_{zx})$-orbitals hybridize 
as represented by the ``odd-parity hybridization term'' $H_{\rm{odd}}$. This term plays an essential role in the following 
results in combination with the LS-coupling term $H_{\rm{LS}}$. 
Note that the odd-parity hybridization term has the opposite sign between layers, 
i.e., $V_{x, \, A}(\bm{k})=-V_{x, \, B}(\bm{k})$ and $V_{y, \, A}(\bm{k})=-V_{y, \, B}(\bm{k})$, 
so that the global mirror symmetry of the bilayer is conserved. 
Two layers are coupled to each other via the interlayer hopping term $H_{\perp}$. 
The interlayer hopping of the quasi-two-dimensional $d_{xy}$-orbital is smaller than those of 
the $d_{yz}$- and $d_{zx}$-orbitals~\cite{Raghu}, namely, $t_{\perp, \, 3} < t_{\perp, \, 1} = t_{\perp, \, 2}$. 
We also take into account the last term $H_{\rm{lp}}$ introducing the folding of the Brillouin zone due to the 
rotation of the RuO$_6$ octahedra. The wave vector is $\bm{Q} = (\pi,\pi)$.

We now adopt the following tight-binding approximation, 
$\varepsilon_{1}(\bm{k}) = -2 t_{3} \cos{k_{x}} - 2 t_{2} \cos{k_{y}}$,  
$\varepsilon_{2}(\bm{k}) = -2 t_{2} \cos{k_{x}} - 2 t_{3} \cos{k_{y}}$,  
$\varepsilon_{3}(\bm{k}) = -2 t_{1} (\cos{k_{x}} + \cos{k_{y}}) - 4 t_{4} \cos{k_{x}} \cos{k_{y}}$,  
$V(\bm{k}) = 4 t_{5} \sin{k_{x}} \sin{k_{y}}$, 
$V_{x, \, A}(\bm{k}) = 2 i \, t_{\rm{odd}} \sin{k_{x}}$, and $V_{y,\, A}(\bm{k}) = 2 i \, t_{\rm{odd}} \sin{k_{y}}$. 
The set of parameters 
$(t_{1}, t_{2}, t_{3}, t_{4}, t_{5}, t_{\rm{odd}}, t_{\perp, \, 1}, t_{\perp, \, 2}, t_{\perp, \, 3}, \lambda, \Delta, g_{\rm{lp}}) 
= (1.0, 1.2, 0.1, 0.25, 0.1, 0.1, 0.6, 0.6, 0.1, 0.3, -0.3, 0.1)$ 
reproduces the electronic structure of Sr$_3$Ru$_2$O$_7$ 
which has been obtained by LDA calculation and consistently observed 
by angle-resolved photo emission spectroscopy (ARPES)~\cite{Tamai} and 
de Haas-van Alphen (dHvA) measurement~\cite{Mercure} (see Fig.~8).  
We adopt this parameter set unless specified. 
We ignore the lattice distortion term $H_{\rm{lp}}$ in Sects. 2 and 3 
since it is not an essential factor of the EO state. 
The effects of lattice distortion on the electronic structure will be elucidated in Sect. 4. 
Thus, we assume $g_{\rm{lp}} = 0$ in Sects. 2 and 3, while we adopt $g_{\rm{lp}} = 0.1$ in Sect. 4.

%\vspace{-5mm}
\begin{figure}[htb]
 \begin{center}
   \includegraphics[width=8cm]{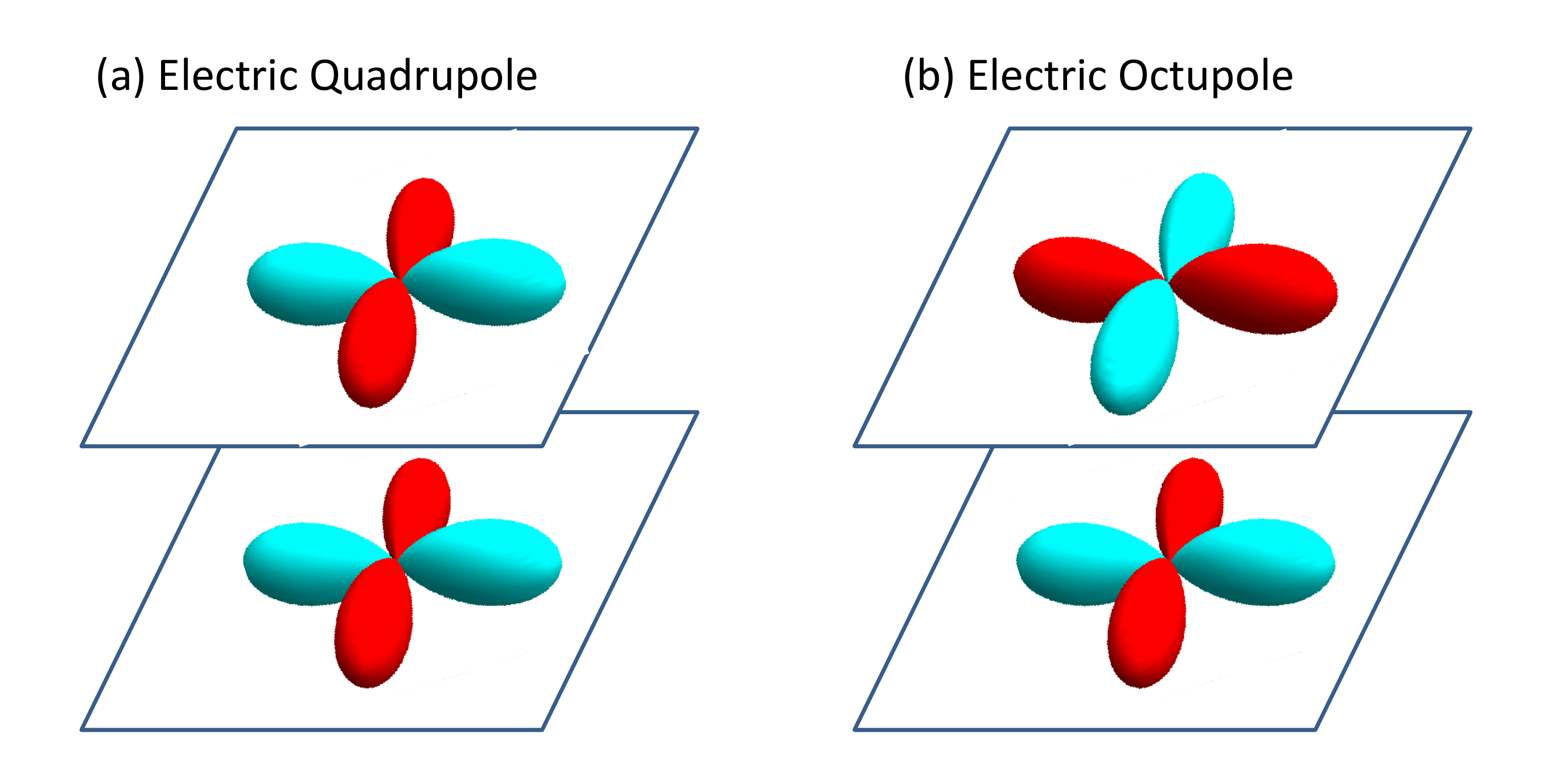}
  \caption{(Color online) Schematic figures of (a) ferro stacking and (b) antiferro stacking of 
the $d$-wave nematic order in the bilayer. The former belongs to the EQ ($O_{x^2-y^2}$) order, while the latter is regarded 
as an  EO ($T_{(x^2-y^2)z}$) order. 
} 
 \end{center}
\end{figure}

Next, we consider the electronic nematic order~\cite{Sr3Ru2O7_review}. 
Instead of studying the many-body effect giving rise to the nematic order, we adopt the mean field terms of $d$-wave nematic order 
caused by the dPI~\cite{Kee-Kim,Yamase-Katanin,Yamase,Adachi} and/or by the orbital order,~\cite{Raghu,Lee-Wu,Ono,Tsuchiizu}
\begin{align}
H_{\rm{dPI}} &= \Delta^{\rm{dPI}} \sum_{\bm{k},s} \varphi_{d}(\bm{k}) 
[c^{\dagger}_{\bm{k}3sA} c_{\bm{k}3sA} \pm c^{\dagger}_{\bm{k}3sB} c_{\bm{k}3sB} ], \label{eq:dPI}
\end{align}
where $\varphi_{d}(\bm{k}) = \cos{k_{x}} - \cos{k_{y}}$ is the $d$-wave form factor, 
and 
\begin{align}
H_{\rm{orb}} = \Delta^{\rm{orb}} \sum_{\bm{k},s} [(&c^{\dagger}_{\bm{k}1sA} c_{\bm{k}1sA} - c^{\dagger}_{\bm{k}2sA} c_{\bm{k}2sA}) \notag \\ 
  \pm (&c^{\dagger}_{\bm{k}1sB} c_{\bm{k}1sB} - c^{\dagger}_{\bm{k}2sB} c_{\bm{k}2sB})]. \label{eq:orb}
\end{align}
The $+$ sign represents the ferro stacking and $-$ sign represents the antiferro stacking between the layers, respectively. 
The former corresponds to the EQ order, while the  EO order is realized in the latter, 
as illustrated in Fig.~1. In the latter case, indeed, the electron charge distribution shows a polarization 
corresponding to the $T_{(x^2-y^2)z}$ octupole whose origin is at the inversion center. 
Note that the inversion center is not on the atom, 
but at the midpoint of two Ru atoms in the bilayer.

\begin{figure}[htb]
 \begin{center}
  \subfigure[Fermi surface]{
   \includegraphics[width=.45\columnwidth]{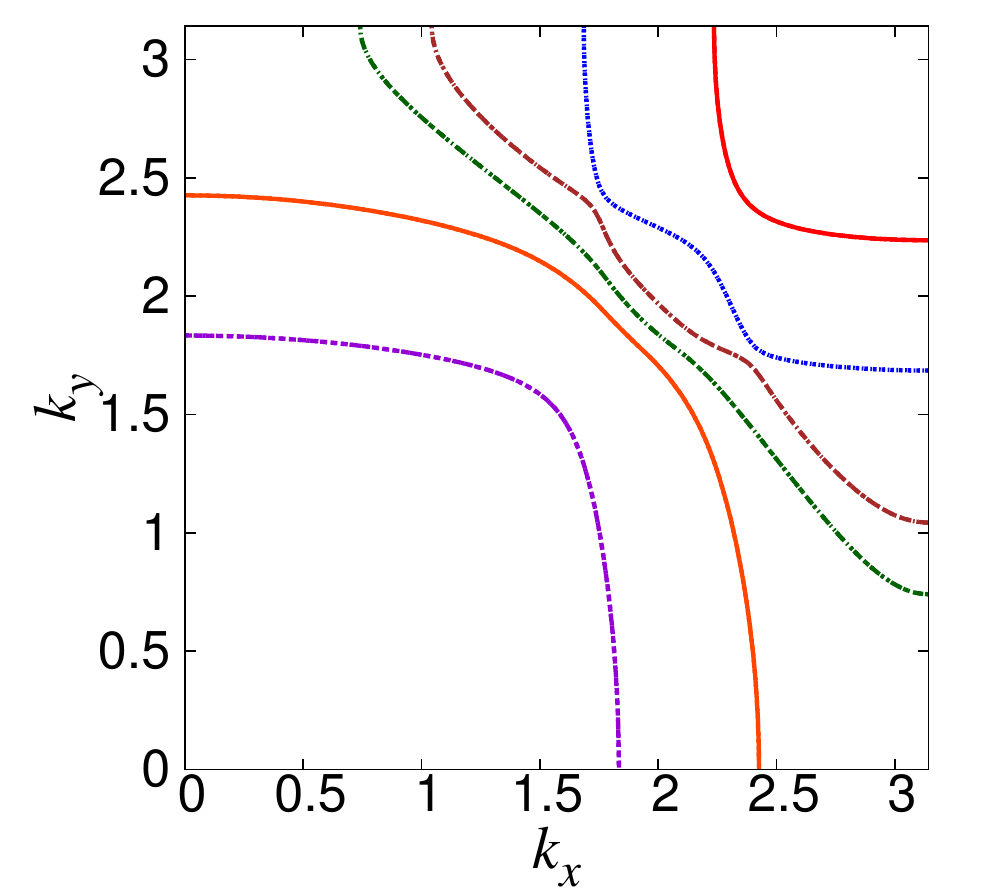}
  }~
  \subfigure[Band structure]{
   \includegraphics[width=.45\columnwidth]{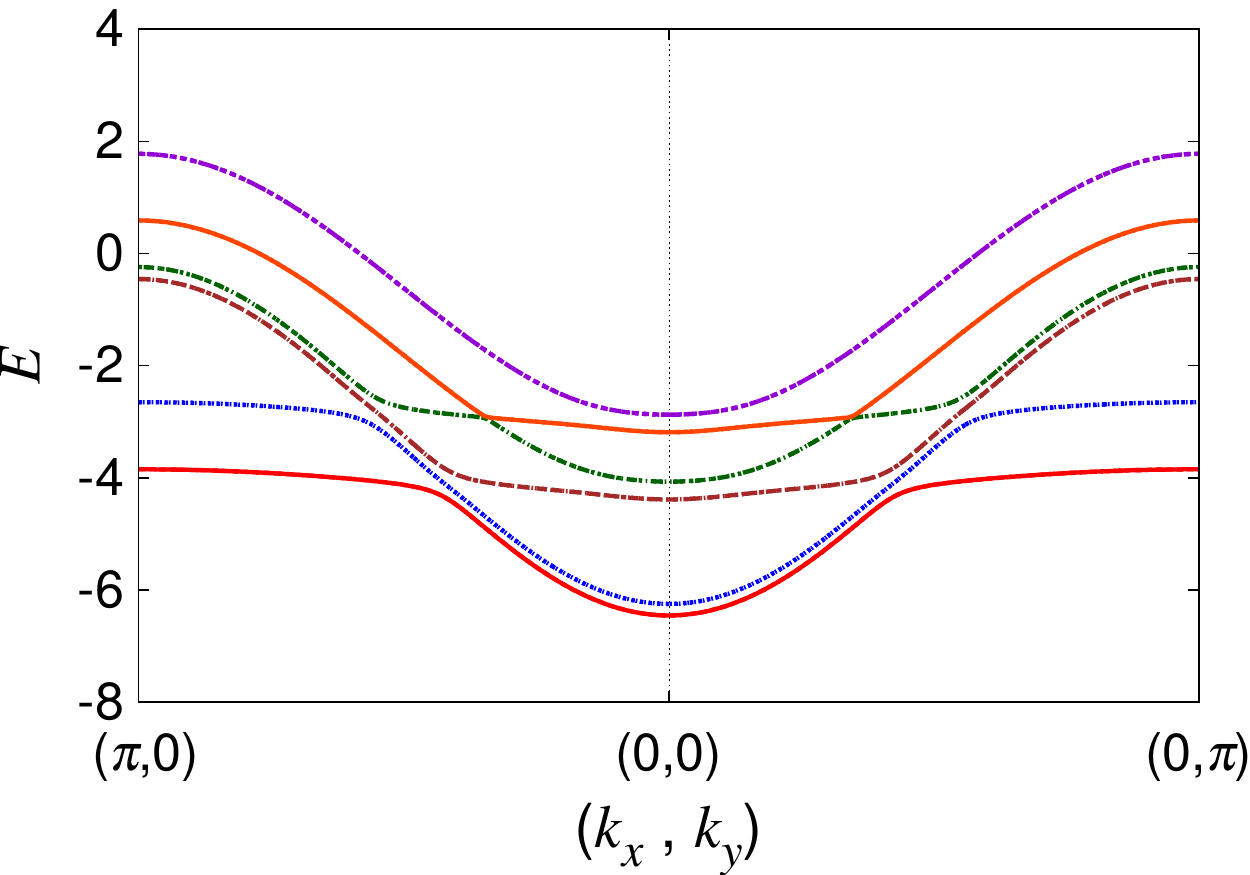}
  }
  \caption{(Color online) Electronic structure in the normal state $(\Delta^{\rm{dPI}}, \Delta^{\rm{orb}}) = (0, 0)$. 
(a) Fermi surface and (b) band structure conserving the $C_{4}$ rotation symmetry and the spin degeneracy. 
} 
  \label{Fig:no_op}
 \end{center}
\end{figure}

The total Hamiltonian is given by 
\begin{equation}
H = H_{\rm{0}} + H_{\rm{dPI}} + H_{\rm{orb}}. \label{eq:H}
\end{equation}
When we ignore the lattice distortion term $H_{\rm{lp}}$, we obtain 12 eigenstates 
for each momentum $\k$. The electronic structure is described by the 6 bands with spin degeneracy. 
For comparison with the EQ and EO states, Fig.~\ref{Fig:no_op} shows the band structure and Fermi surface in the normal state, 
where $(\Delta^{\rm{dPI}} , \Delta^{\rm{orb}}) = (0, 0)$. 
The 1$^{\rm st}$, 2$^{\rm nd}$, 5$^{\rm th}$, and 6$^{\rm th}$ Fermi surfaces mainly consist of the $d_{yz}$- and $d_{zx}$-orbitals, while 
the 3$^{\rm rd}$ and 4$^{\rm th}$ Fermi surfaces have the $d_{xy}$-orbital characteristic.

\subsection{Ferro stacking: electric quadrupole order}

%First, we study the EQ state. 
Figures~3(a) and 3(b) show the Fermi surfaces and band structure when the EQ order is caused by the dPI, 
$(\Delta^{\rm{dPI}}, \Delta^{\rm{orb}}) = (1, 0)$. On the other hand, Figs.~3(c) and 3(d) are obtained for the orbital ordered state, 
where $(\Delta^{\rm{dPI}}, \Delta^{\rm{orb}}) = (0, 1)$.  
The large amplitudes of the order parameters $\Delta^{\rm{dPI}}$ and $\Delta^{\rm{orb}}$ are assumed 
so as to emphasize the effect of the EQ order. 
We see that the Fermi surfaces are spontaneously deformed and the rotation symmetry is reduced from $C_{4}$ to $C_{2}$. 
Indeed, this is the consequence of the EQ order, which has been discussed 
in the literature~\cite{Sr3Ru2O7_review,Kee-Kim,Yamase-Katanin,Yamase,Adachi,Raghu,Lee-Wu}. 
We stress that each band holds the spin degeneracy in contrast to the  EO state, which we study in the next subsection. 
\begin{figure}[htb]
  \begin{center}
    \subfigure[Fermi surface]{
      \includegraphics[width=.45\columnwidth]{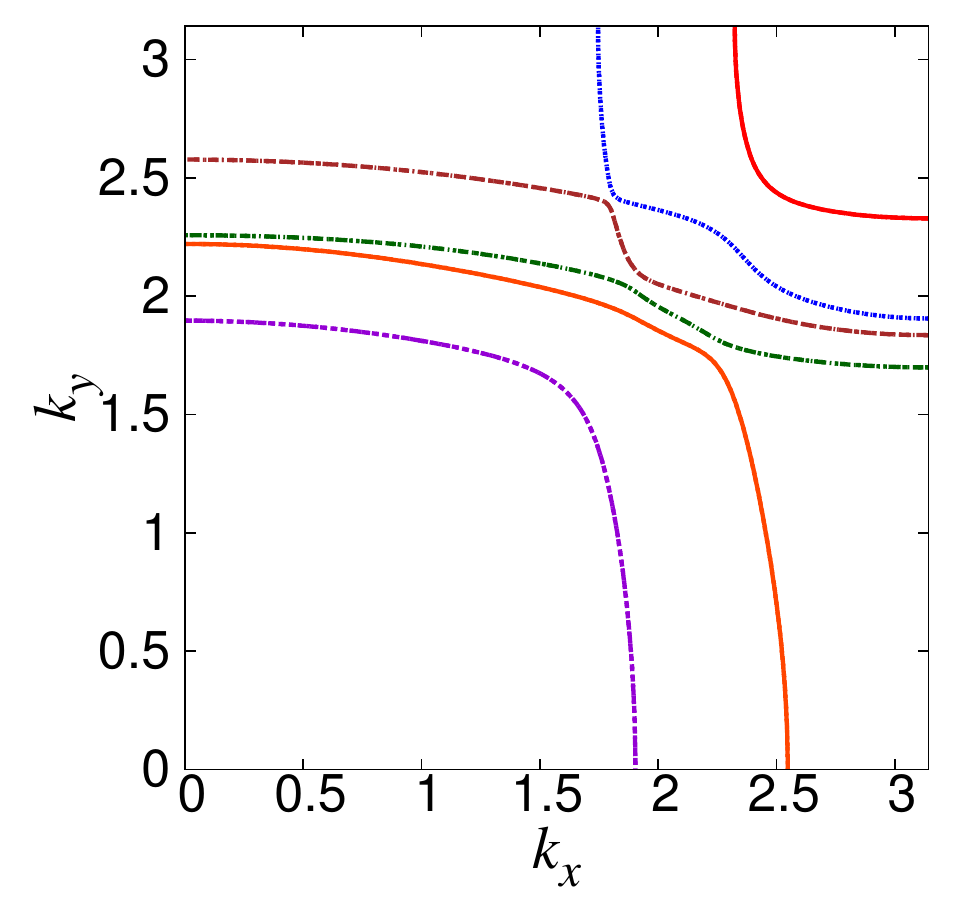}
    }~
    \subfigure[Band structure]{
      \includegraphics[width=.45\columnwidth]{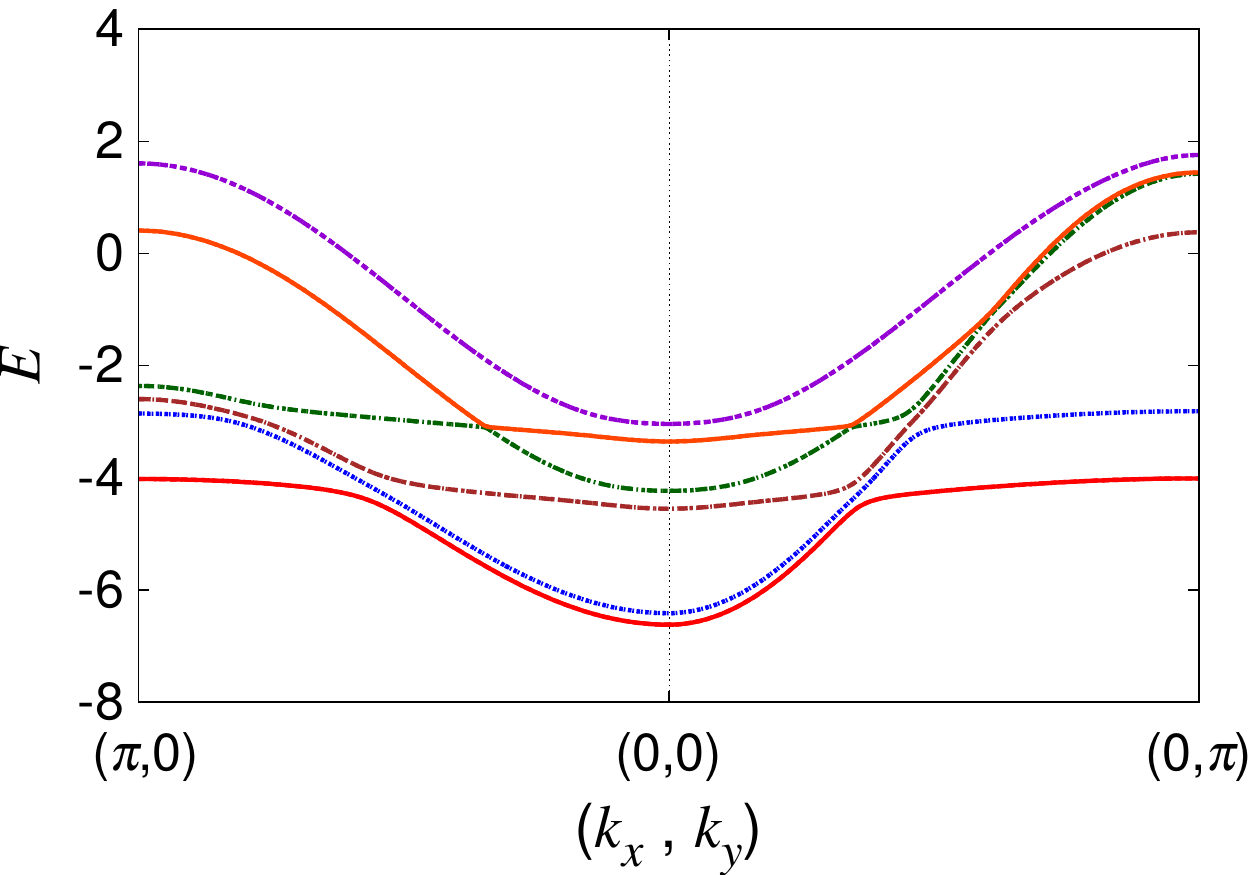}
    } \\
    \subfigure[Fermi surface]{
      \includegraphics[width=.45\columnwidth]{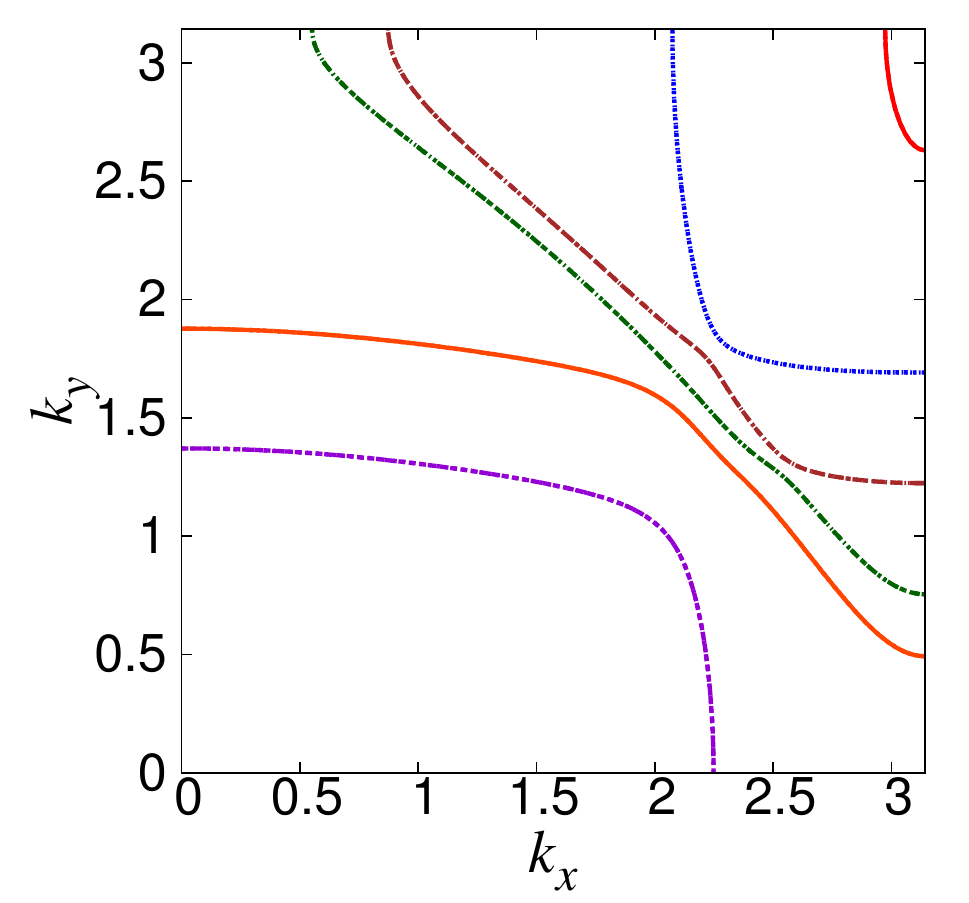}
    }~
    \subfigure[Band structure]{
      \includegraphics[width=.45\columnwidth]{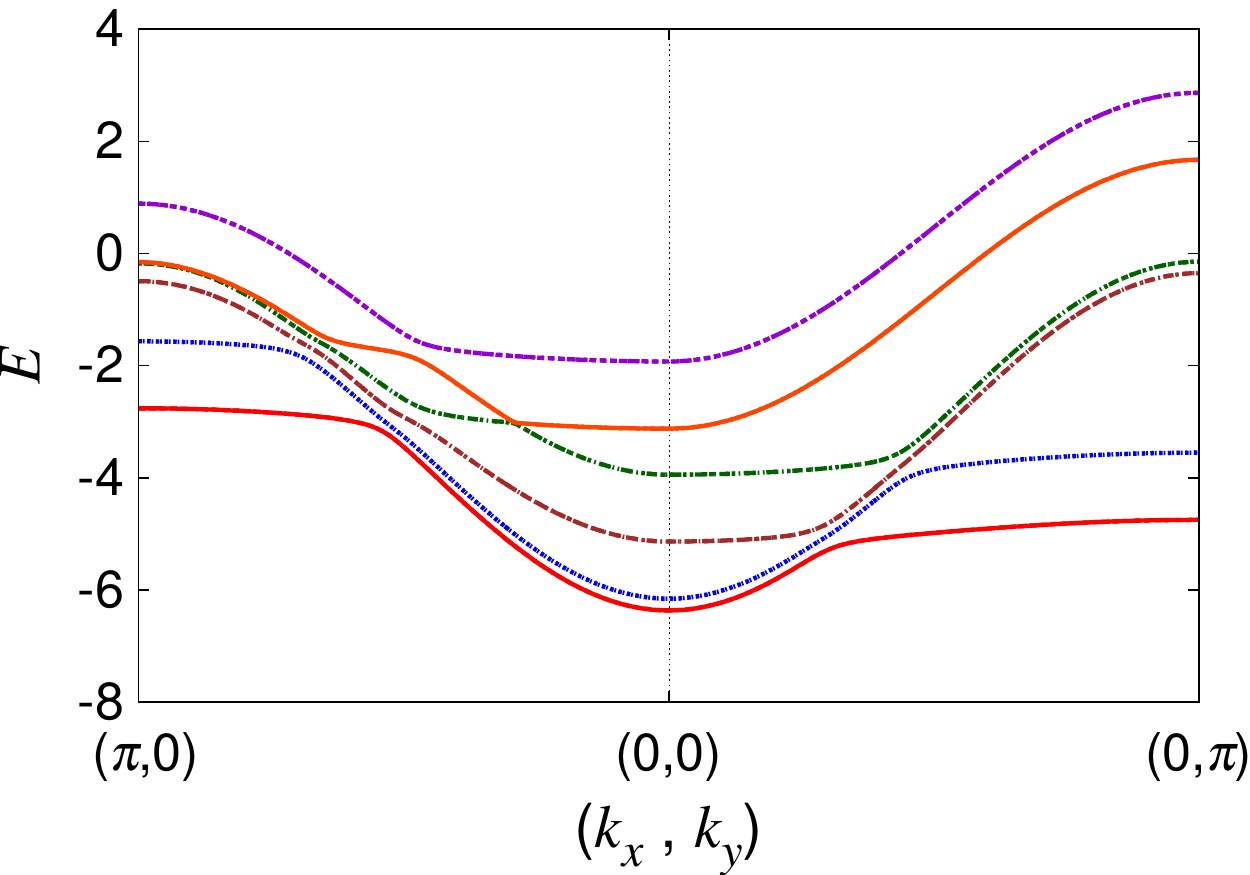}
    }
    \caption{(Color online) Electronic structure in the EQ state. 
(a) and (b) are obtained for the dPI-dominated quadrupole order, i.e., $(\Delta^{\rm{dPI}}, \Delta^{\rm{orb}}) = (1, 0)$. 
(c) and (d) are obtained for the orbital-order-dominated quadrupole order, i.e., $(\Delta^{\rm{dPI}}, \Delta^{\rm{orb}}) = (0, 1)$. 
(a) and (c) show the spontaneous rotation symmetry breaking in the Fermi surfaces. 
(b) and (d) show the band structure holding the spin degeneracy. 
} 
    \label{Fig:F_op}
  \end{center}
\end{figure}

Figure~3 enables us to distinguish the dPI-dominated EQ order from the orbital-order-dominated one. 
The Fermi surfaces of the 3$^{\rm rd}$ and 4$^{\rm th}$ bands are markedly distorted in the dPI-dominated case [Fig.~3(a)] 
because these bands mainly consist of the $d_{xy}$-orbital. 
On the other hand, the 1$^{\rm st}$, 2$^{\rm nd}$, 5$^{\rm th}$, and 6$^{\rm th}$ bands are considerably affected by the 
orbital order of ($d_{yz}$, $d_{zx}$)-orbitals. Thus, the mechanism of quadrupole order manifests in the band structure.

\subsection{Antiferro stacking: electric octupole order}

\begin{figure}[htb]
 \begin{center}
  \subfigure[Fermi surface]{
   \includegraphics[width=.48\columnwidth]{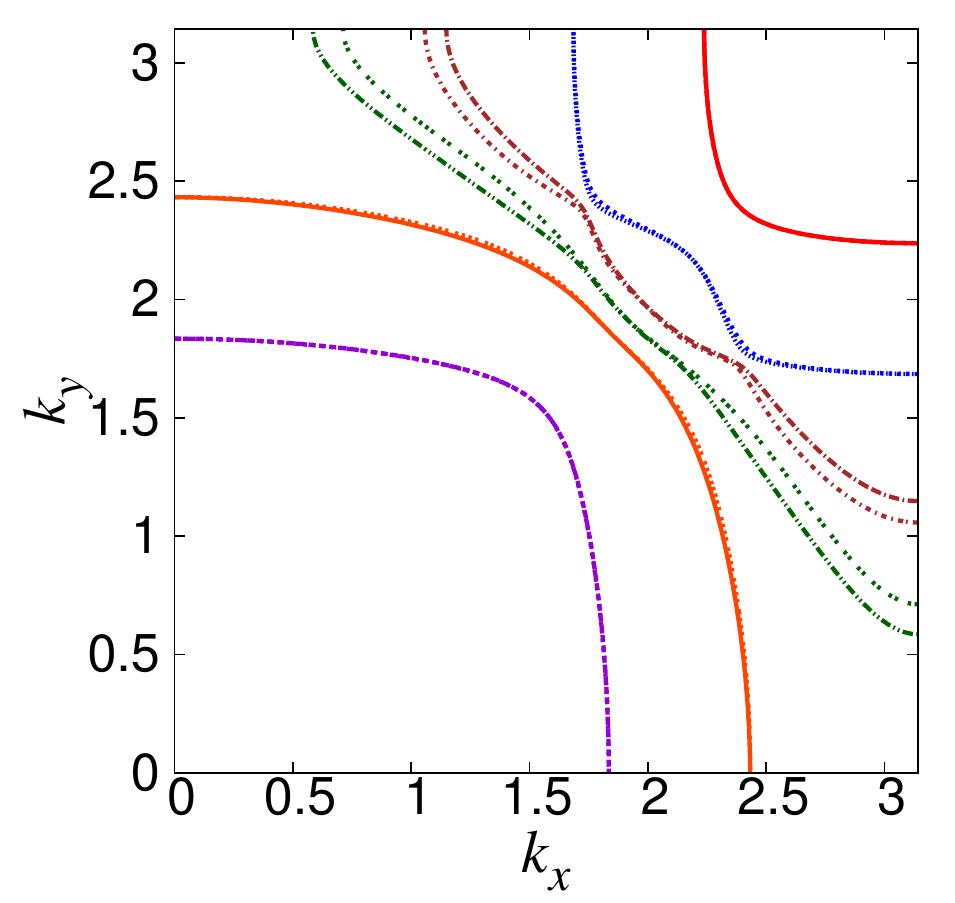}
  }~
  \subfigure[Fermi surface]{
   \includegraphics[width=.48\columnwidth]{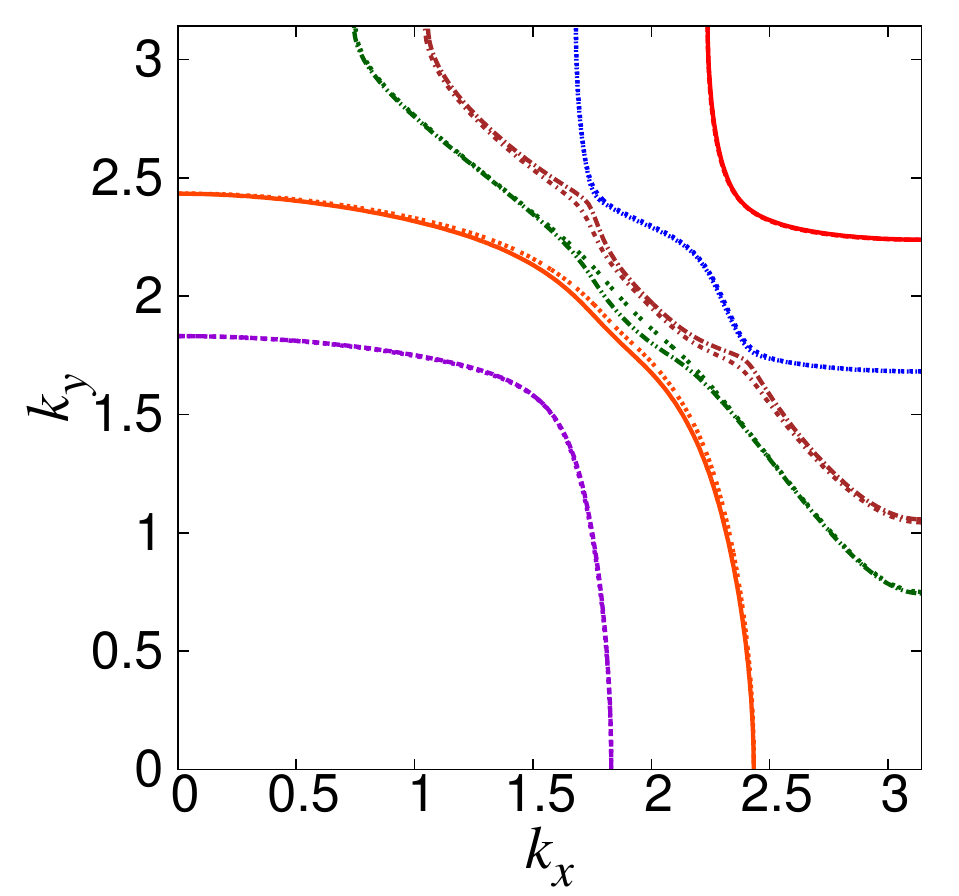}
  }
  \caption{(Color online) Fermi surfaces in the (a) dPI-dominated  EO state, $(\Delta^{\rm{dPI}}, \Delta^{\rm{orb}}) = (0.08, 0)$, 
and (b) orbital-order-dominated  EO state, 
$(\Delta^{\rm{dPI}}, \Delta^{\rm{orb}}) = (0, 0.08)$. The $C_4$ rotation symmetry is conserved, but the spin degeneracy is lifted.  
}
  \label{Fig:AF_FS_op}
 \end{center}
\end{figure}

Next, we address the consequences of the  EO order. 
Figures~\ref{Fig:AF_FS_op}(a) and 4(b) show the Fermi surfaces in the  EO states induced by the dPI and orbital order, 
respectively. In both cases, we see the spin splitting of Fermi surfaces as in the noncentrosymmetric metal~\cite{NCSC}. 
Indeed, the space inversion symmetry is spontaneously broken in the odd-parity multipole state. 

The odd-parity  EO order gives rise to the spin splitting in the Fermi surfaces in combination with 
the LS coupling and odd-parity hybridization terms. 
In other words, the spin splitting does not occur when either $\lambda$ or $t_{\rm odd}$ is zero. 
Because the odd-parity hybridization term results from the local violation of inversion symmetry in the bilayer crystal structure~\cite{a}, 
it is concluded that the local noncentrosymmetricity in the crystal structure plays an essential role 
in the electronic structure in the  EO state. 
Thus, the elaborate model adopted in this paper is necessary for the description of the  EO state. 
Although the antiferro stacking of the $d$-wave nematic order has been studied~\cite{Puetter_bilayer,Yamase_bilayer}, 
some essential factors have been neglected.

Figure 4 shows the pronounced spin splitting in the 3$^{\rm rd}$ and 4$^{\rm th}$ bands. This is because the spin-orbit coupling 
generally competes with the single-particle interlayer hopping in multilayer systems~\cite{f}. 
As we mentioned in Sect.~2.1, the interlayer hopping is small for the $d_{xy}$-orbital because of its two-dimensional character; 
therefore, the spin-orbit coupling significantly affects the $d_{xy}$-orbital. Thus, the 3$^{\rm rd}$ and 4$^{\rm th}$ bands, 
which mainly consist of the $d_{xy}$-orbital, show a large spin splitting. 
This explanation is confirmed in Fig.~5, which shows the average spin splitting energy on the Fermi surface 
defined for the $j$-th band as 
\begin{align}
\left< \Delta E \right>_j = \rho_{j}^{-1} \sum_\k \left(E_{2j}(\k) - E_{2j-1}(\k)\right) \left[\delta(E_{2j}(\k))+\delta(E_{2j-1}(\k))\right], 
\label{splitting_FS}
\end{align}
where $E_{i}(\k)$ is the $i$-th eigenvalue of the total Hamiltonian, 
and $\rho_{j} = \sum_\k \left[\delta(E_{2j}(\k))+\delta(E_{2j-1}(\k))\right]$ is the density of states of the $j$-th band.
Note that $E_{2j}(\k) \ne E_{2j-1}(\k)$ except for the time-reversal invariant momentum in the Brillouin zone. 
Figure~5(a) shows the decrease in the spin splitting energy 
in the 3$^{\rm rd}$ and 4$^{\rm th}$ bands with increasing interlayer hopping of the $d_{xy}$-orbital, $t_{\perp, \, 3}$. 
The other bands are almost unchanged by the increase in $t_{\perp, \, 3}$. 
On the other hand, the spin splitting energy in the 1$^{\rm st}$, 2$^{\rm nd}$,  5$^{\rm th}$, and 6$^{\rm th}$ bands decreases with 
increasing $t_{\perp, \, 1} = t_{\perp, \, 2}$ when the EO order is induced by the orbital order [Fig.~5(b)]. 
When the interlayer hopping is orbital-independent, $t_{\perp, \, 1} = t_{\perp, \, 2} = t_{\perp, \, 3}$, the orbital order induces a 
larger spin splitting in these bands than in the 3$^{\rm rd}$ and 4$^{\rm th}$ bands. However, the realistic parameter set 
$(t_{\perp, \, 1}, t_{\perp, \, 2}, t_{\perp, \, 3}) = (0.6, 0.6, 0.1)$ leads to a substantial spin splitting in the 
$d_{xy}$-orbital-dominated bands even when the EO order is induced by the ($d_{yz}$, $d_{zx}$)-orbital order. 
Table I shows a summary of the average spin splitting energy in the dPI-dominated and orbital-order-dominated  EO states. 

\begin{figure}[htb]
 \begin{center}
   \includegraphics[width=6.5cm]{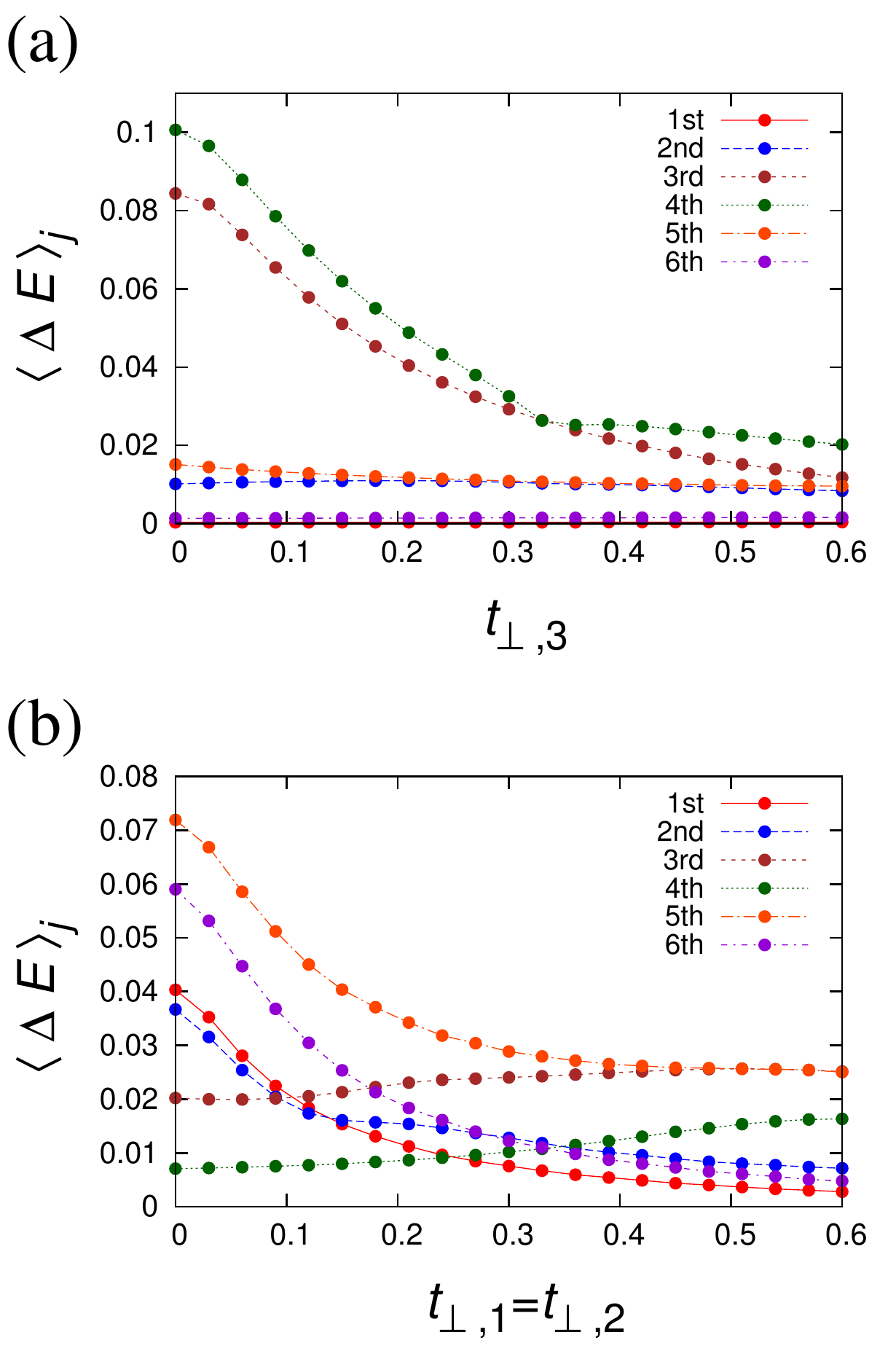}
  \caption{(Color online) Average spin splitting energy on the Fermi surfaces, $\left< \Delta E \right>_j$, 
(a) as a function of $t_{\perp, \, 3}$ and (b) as a function of $t_{\perp, \, 1} = t_{\perp, \, 2}$. 
The definition of $\left< \Delta E \right>_j$ has been given in Eq.~(\ref{splitting_FS}). 
We assume $(\Delta^{\rm{dPI}}, \Delta^{\rm{orb}}) = (0.08, 0)$ (dPI-dominated EO order) 
and $t_{\perp, \, 1} = t_{\perp, \, 2} =0.6$ in Fig.~5(a), 
and $(\Delta^{\rm{dPI}}, \Delta^{\rm{orb}}) = (0, 0.08)$ (orbital-order-dominated EO order) and $t_{\perp, \, 3}=0.1$ in Fig.~5(b). 
}
 \end{center}
\end{figure}

\begin{table}[htbp]
\begin{center}
{\renewcommand\arraystretch{1.2}
\begin{tabular}{c||c|c}
& dPI  & Orbital Order 
\\
\hline
\hline
$\left< \Delta E \right>_1$ & 0.00033 & 0.00278
\\
\hline
$\left< \Delta E \right>_2$ & 0.01074 & 0.00716
\\
\hline
$\left< \Delta E \right>_3$ & 0.06275 & 0.02511
\\
\hline
$\left< \Delta E \right>_4$ & 0.07558 & 0.01633
\\
\hline
$\left< \Delta E \right>_5$ & 0.01311 & 0.02500
\\
\hline
$\left< \Delta E \right>_6$ & 0.00136 & 0.00477
\\
\hline
\end{tabular}
%\end{eqnarray}
}
\end{center}
\caption{
Band dependence of the average spin splitting energy, $\left< \Delta E \right>_j$. 
The middle column shows the results for $(\Delta^{\rm{dPI}}, \Delta^{\rm{orb}}) = (0.08, 0)$ (dPI-dominated  EO order), 
while the right column assumes  $(\Delta^{\rm{dPI}}, \Delta^{\rm{orb}}) = (0, 0.08)$ (orbital-order-dominated EO order). 
}
\label{table1}
%\vspace*{5mm}
\end{table}

We see a marked difference in the momentum dependence of spin splitting between the dPI-dominated 
and orbital-order-dominated  EO states. 
To reveal the difference, Fig.~6 shows the momentum dependence of the spin splitting energy in the 4$^{\rm th}$ band. 
Generally speaking, the effect of spin-orbit coupling is enhanced at $|k_x| = |k_y|$, 
where the $t_{2g}$-orbitals are nearly degenerate. 
Indeed, the spin splitting is enhanced at $|k_x| = |k_y|$ in the orbital-order-dominated  EO state [Fig.~6(b)]. 
On the other hand, the spin splitting vanishes at $|k_x| = |k_y|$ in the dPI-dominated  EO state [Fig.~6(a)], 
because of the $d$-wave form factor $\varphi_{d}(\k)$. 
Thus, the origin of the  EO order may be identified by the measurement of spin-split bands using ARPES 
or by performing quantum oscillation experiments.

\begin{figure}[htb]
 \begin{center}
\hspace*{-10mm}
   \includegraphics[width=10.5cm]{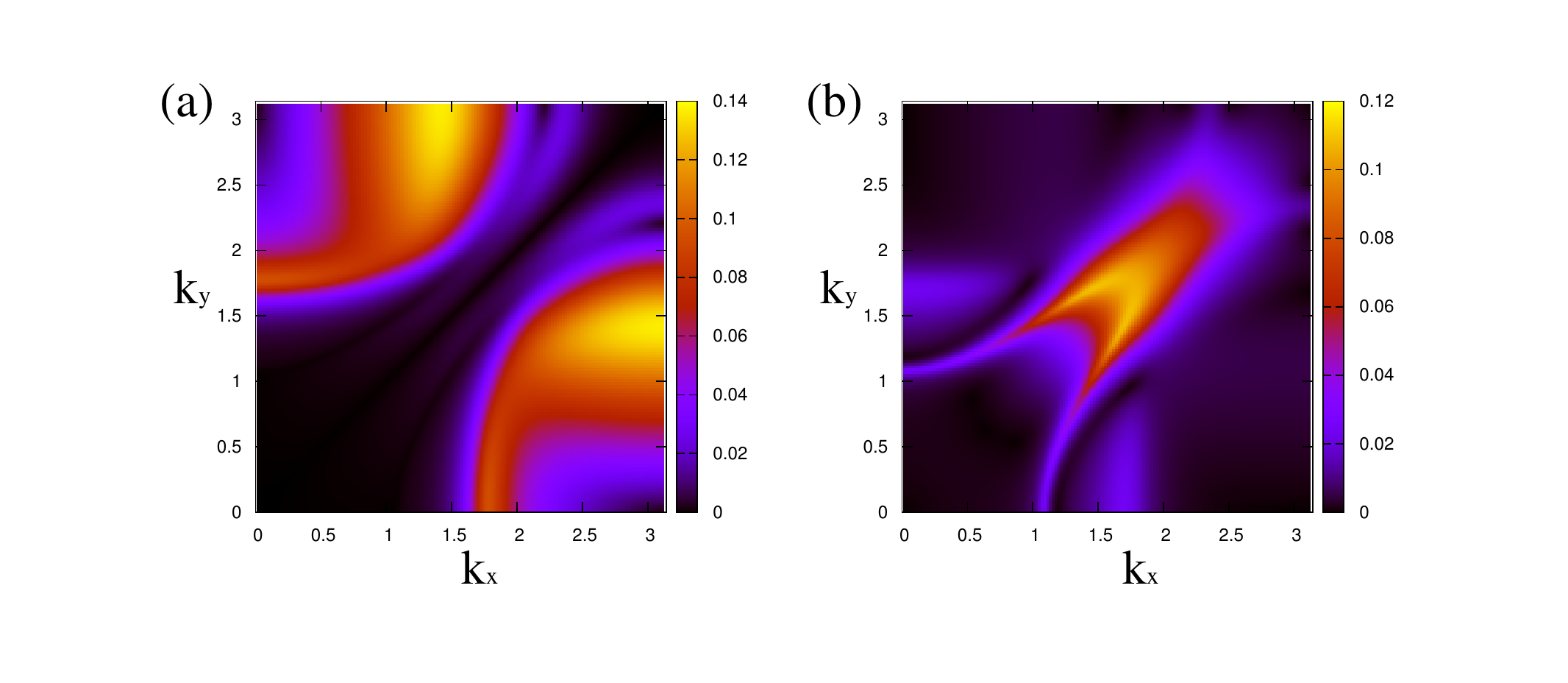}
  \caption{(Color online) 
Momentum dependence of the spin splitting energy in the 4$^{\rm th}$ band, $E_{8}(\k) - E_{7}(\k)$. 
(a) dPI-dominated  EO state, $(\Delta^{\rm{dPI}}, \Delta^{\rm{orb}}) = (0.08, 0)$ 
and 
(b) orbital-order-dominated  EO state, $(\Delta^{\rm{dPI}}, \Delta^{\rm{orb}}) = (0, 0.08)$.
}
 \end{center}
\end{figure}

\section{Spin Texture and Spin Nematic Order}

\subsection{Spin texture in the momentum space} 

In this section, we focus on the  EO state and illustrate the spin texture in the momentum space. 
Since the spin degeneracy in the band is lifted owing to the spontaneous inversion symmetry breaking, the split bands show the spin texture. 
As we will show in Sect.~5, the spin texture can be examined by experiments in the magnetic field. 
For Sr$_3$Ru$_2$O$_7$, the nematic order occurs in the magnetic field; therefore, the experiment 
is naturally carried out in the magnetic field.

The spin texture is characterized by the ``g-vector'' defined as 
\begin{eqnarray}
\label{g-vector}
&& \hspace{-10mm}
\g_j({\bm k}) = \frac{1}{2} \left(E_{2j}({\bm k})-E_{2j-1}({\bm k})\right) \, \frac{\SS_{2j}^{\rm \ av}({\bm k})}{\left|\SS_{2j}^{\rm \ av}({\bm k})\right|},  
\end{eqnarray} 
where $\SS_{i}^{\rm \ av}({\bm k}) = \langle \sum_{ml}\sum_{ss'} 
{\bm \sigma}^{ss'} c_{{\bm k} \, m s l}^{\dag}c_{{\bm k} \, m s' l} \rangle_i$ is the expected value of the spin for the $i$-th eigenstate. 
The direction of the g-vector represents the spin texture in the $j$-th band and its amplitude indicates the 
spin splitting energy. 
We confirmed $\SS_{2j}^{\rm \ av}({\bm k}) \simeq - \SS_{2j-1}^{\rm \ av}({\bm k})$ and $\SS_{i}^{\rm \ av}({\bm k}) = - \SS_{i}^{\rm \ av}(-{\bm k})$.  
Therefore, the band structure is approximately described by the multi-band model for the noncentrosymmetric metal~\cite{a,Nakamura_STO}, 
\begin{eqnarray}
\label{band-basis}
%&& \hspace*{-10mm}  
%H_{\rm 0} = H_{\rm band} + H_{\rm ASOC}, 
%\\ 
&& \hspace*{-12mm}  
H_{\rm eff} =\sum_{j=1}^6 \sum_{{\bm k}, s, s'} \left[ \xi_{j}({\bm k}) \ \sigma_0^{ss'} + 
\g_{j}({\bm k}) \cdot {\bm \sigma}^{ss'} \right] \ a_{{\bm k} \, j s}^{\dag} \ a_{{\bm k} \, j s'}, 
%\nonumber \\ &&
\end{eqnarray}
where $\xi_{j}({\bm k}) = \left(E_{2j}({\bm k}) + E_{2j-1}({\bm k})\right)/2$. 
Thus, the spin splitting due to the odd-parity  EO order is described by introducing 
the effective antisymmetric spin-orbit coupling~\cite{Comment1}.

\begin{figure}[htbp]
 \begin{center}
   \includegraphics[width=8.2cm]{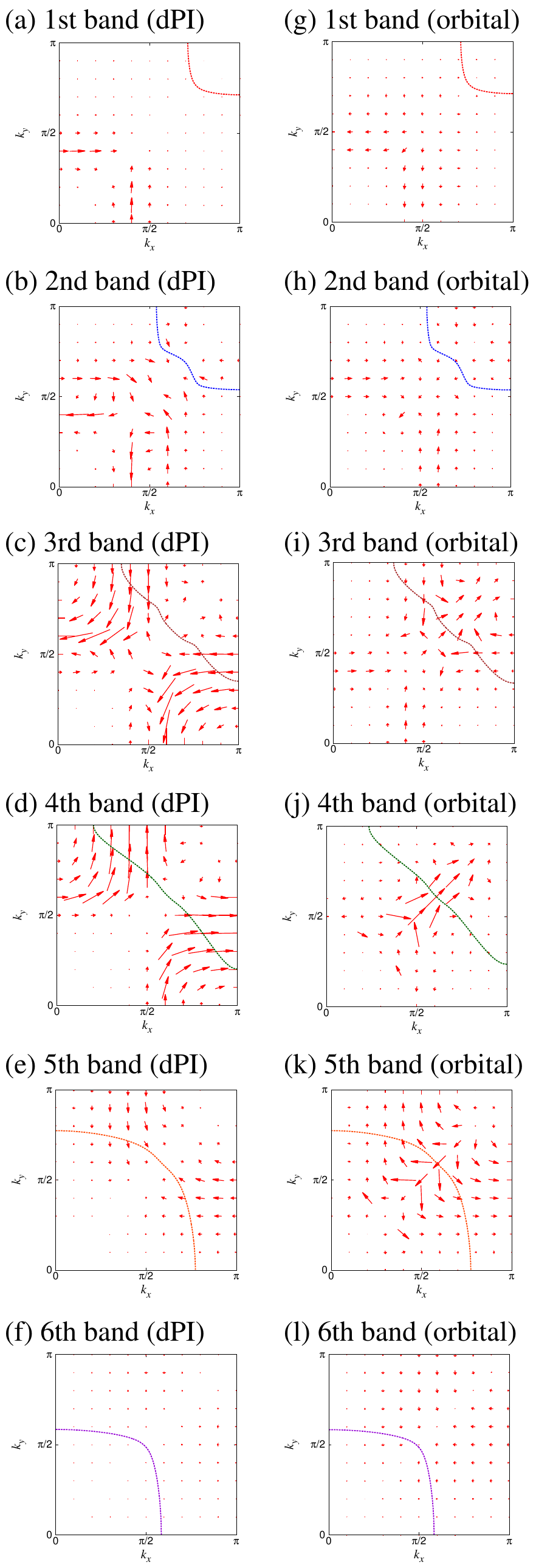}
  \caption{(Color online) g-vector defined in Eq.~(\ref{g-vector}). The momentum dependence in the quarter Brillouin zone is 
shown for each band. The Fermi surfaces are also drawn by dotted lines. 
We assume the dPI-dominated  EO state $(\Delta^{\rm{dPI}}, \Delta^{\rm{orb}}) = (0.08, 0)$ in (a)-(f) 
and the orbital-order-dominated  EO state $(\Delta^{\rm{dPI}}, \Delta^{\rm{orb}}) = (0, 0.08)$ in (g-l).
}
 \end{center}
\end{figure}
Figure~7 shows the momentum dependence of g-vectors in the quarter Brillouin zone. 
Since we find that the $z$-component is zero, the in-plane component of the g-vector is indicated by arrows. 
It is shown that all the g-vectors follow the symmetry $k_y \, \hat{x} + k_x \, \hat{y}$, although we see a 
distinct difference between the dPI-dominated and orbital-order-dominated EO orders.

The $k_y \, \hat{x} + k_x \, \hat{y}$ symmetry of the spin texture results from the $D_{2d}$ point group symmetry of the  EO state. 
The generators of the $D_{2d}$ point group are 
(1) $\pi/2$ rotation for the $z$-axis + mirror reflection for the $xy$-plane and (2) mirror reflection for the $xz$-plane. 
The g-vector compatible with this symmetry operation is obtained as 
$\g({\bm k}) \, =\, k_y \, \hat{x} \, + \, k_x \, \hat{y} \, + \, \gamma \, k_x \, k_y \, k_z \, \hat{z} $. 
Indeed, the spin texture in Fig.~7 belongs to this symmetry class, although the $z$-component disappears 
since we consider the two-dimensional system.

\subsection{$P$-wave spin nematic order}

As we discussed above, the spin splitting due to the  EO order is regarded as a manifestation of the 
antisymmetric spin-orbit coupling. Inversely, we can consider that the EO state is accompanied by 
the $p$-wave spin nematic order through the spin-orbit coupling. 
The induced nematic order parameter has the $p_y s_x + p_x s_y$ symmetry, which is defined as  
\begin{eqnarray}
\label{spin_nematic_order}
&& \hspace*{-8mm}  
\Delta_{\rm psn} = \frac{1}{2} \sum_{l=A,B} \sum_{m=1}^{3} \sum_{\k, s, s'} (\sin k_y \, \sigma_x^{ss'} + \sin k_x \, \sigma_y^{ss'})
\left< c_{\bm{k}msl}^{\dagger} \, c_{\bm{k}ms'l}  \right>. 
\nonumber \\ &&
\end{eqnarray}
We obtain a finite $p$-wave spin nematic order parameter in the  EO state; for example, $\Delta_{\rm psn} = -0.00320$ 
for $(\Delta^{\rm{dPI}}, \Delta^{\rm{orb}}) = (0.08, 0)$ and $\Delta_{\rm psn} = 0.00133$ for 
$(\Delta^{\rm{dPI}}, \Delta^{\rm{orb}}) = (0, 0.08)$. 
Note that $\Delta_{\rm psn}$ vanishes when either the LS coupling $\lambda$ or the odd-parity hybridization $t_{\rm odd}$ is absent.
Thus, the $p$-wave spin nematic order is induced in the EO state by the cooperative roles of the spin-orbit coupling 
and broken local inversion symmetry.  
This result is in parallel to the observation in the magnetic quadrupole state, where  
the $p$-wave charge nematic order is induced~\cite{Magnetic_Quadrupole}. 
Generally speaking, the odd-parity multipole order is accompanied by an odd-parity nematic order through the spin-orbit coupling.

It has been shown that not only the dPI and orbital order but also the $p$-wave spin nematic order can be induced 
by Coulomb interactions~\cite{Yoshioka}. 
This means that the induced $p$-wave spin nematic order further stabilizes the dPI and orbital order 
through the electron correlation effects. 
We will examine the thermodynamical stability of the EQ and EO states 
by taking into account both Coulomb interactions and spin-orbit coupling in a future study. 
In analogy to the observation in this work, 
the antiferro stacking of the $p$-wave spin nematic order gives rise to the ferro dPI order through the spin-orbit coupling, 
when the former is the leading instability. 
This is another scenario for the nematic order in Sr$_3$Ru$_2$O$_7$.

\section{Effect of Lattice Distortion}

The Brillouin zone of Sr$_3$Ru$_2$O$_7$ is folded owing to the lattice distortion 
arising from the rotation of the RuO$_6$ octahedra. 
Here, we investigate the effect of lattice distortion so that the  EO order would be examined by experiments. 
Although a sophisticated model for the lattice distortion potential was obtained~\cite{Fischer}, 
we adopt the last term of Eq.~(\ref{H_0}) for simplicity, as in Ref.~\citen{Puetter_distortion}. 
Our aim to clarify the electronic structure in the folded Brillouin zone is satisfied by our model. 
Figure~8 shows the Fermi surfaces in the normal state for $g_{\rm lp}=0.1$, which reproduces the results of the ARPES~\cite{Tamai} 
and dHvA measurements~\cite{Mercure} as well as the band structure calculation,~\cite{Tamai} although the small electron pocket 
around the $\Gamma$ point arising from the $d_{x^2-y^2}$-orbital~\cite{Tamai} is not reproduced.

\begin{figure}[htb]
 \begin{center}
%\hspace*{-2mm}
   \includegraphics[width=6.5cm]{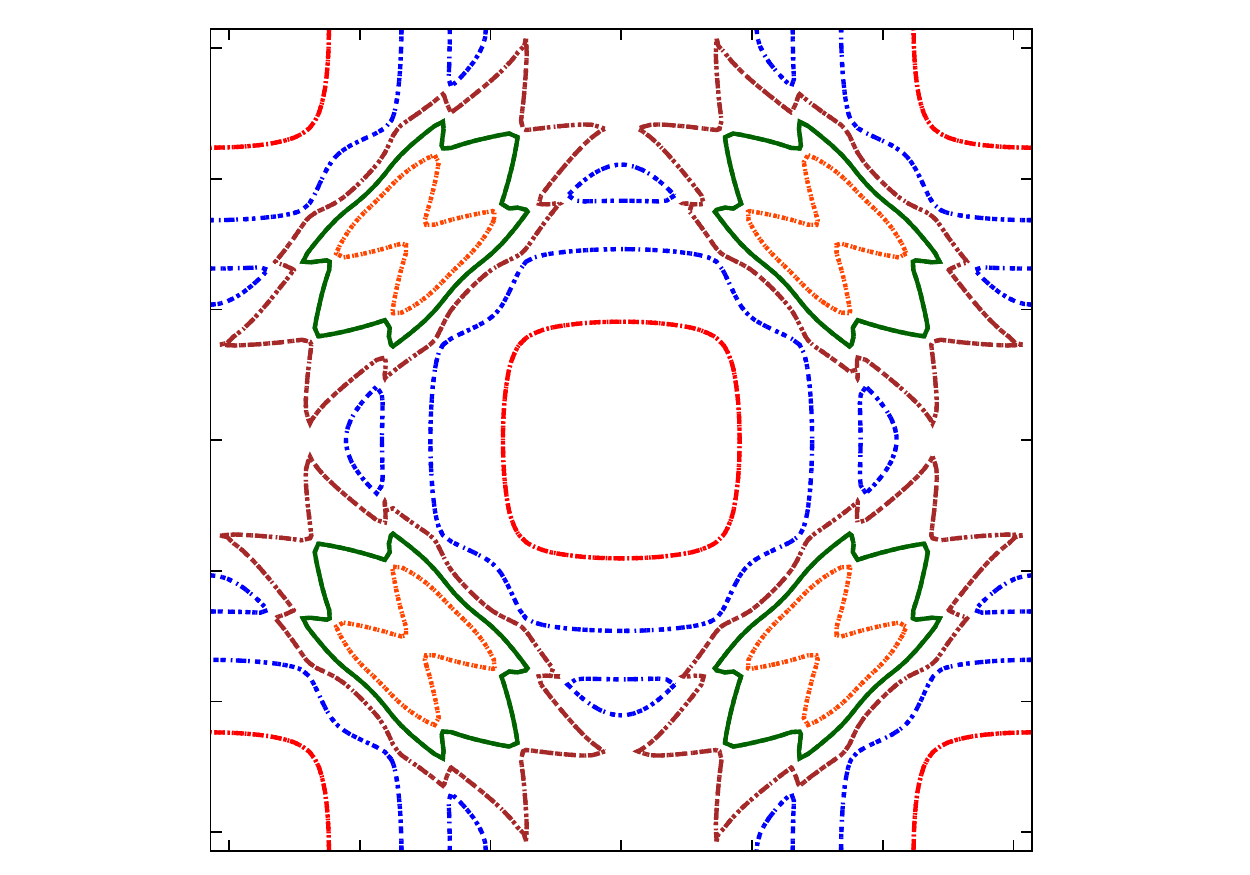}
  \caption{(Color online) 
Fermi surfaces folded by the lattice distortion. 
We assume $g_{\rm lp} =0.1$ and $(\Delta^{\rm{dPI}}, \Delta^{\rm{orb}}) = (0, 0)$. 
}
 \end{center}
\end{figure}

\begin{table}[htbp]
\begin{center}
{\renewcommand\arraystretch{1.2}
\begin{tabular}{c||c|c}
& dPI  & Orbital Order 
\\
\hline
$\left< \Delta E \right>_7$ & 0.00115 & 0.00288
\\
\hline
$\left< \Delta E \right>_8$ & 0.03705 & 0.00787
\\
\hline
$\left< \Delta E \right>_9$ & 0.05489 & 0.01518
\\
\hline
$\left< \Delta E \right>_{10}$ & 0.03408 & 0.03012
\\
\hline
$\left< \Delta E \right>_{11}$ & 0.00780 & 0.03002
\\
\hline
\end{tabular}
}
\end{center}
\caption{
Average spin splitting energy for $g_{\rm lp} =0.1$. 
Owing to the lattice distortion, the Brillouin zone is folded and the band index runs from 1 to 12. 
We show the results for the 7$^{\rm th}$ to 11$^{\rm th}$ bands, which cross the Fermi level. 
The middle column shows the results for $(\Delta^{\rm{dPI}}, \Delta^{\rm{orb}}) = (0.08, 0)$ 
(dPI-dominated  EO order), 
while the right column assumes  $(\Delta^{\rm{dPI}}, \Delta^{\rm{orb}}) = (0, 0.08)$ 
(orbital-order-dominated  EO order). 
}
\label{table2}
%\vspace*{5mm}
\end{table}

\begin{figure}[htb]
 \begin{center}
\hspace*{-2mm}
   \includegraphics[width=9.0cm]{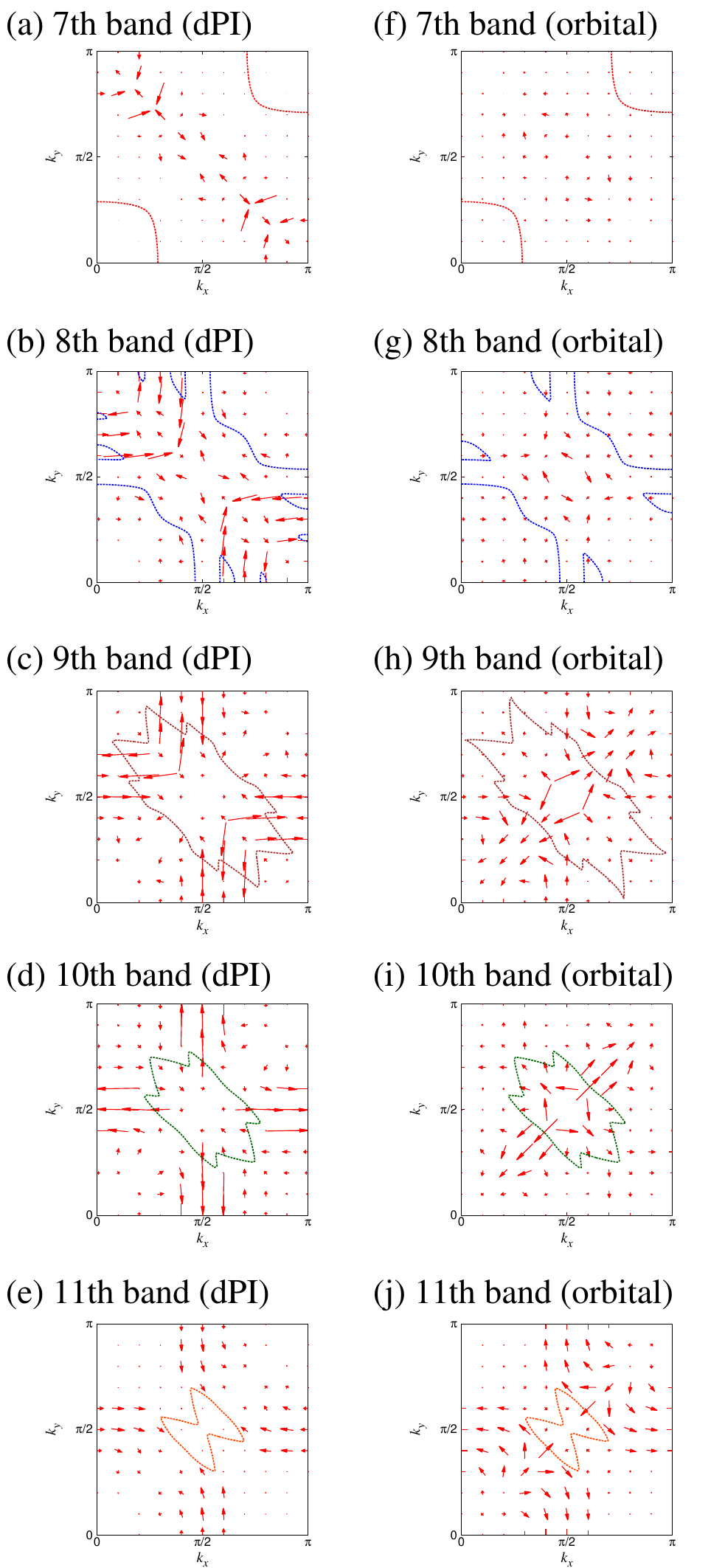}
  \caption{(Color online) 
Momentum dependence of g-vectors in the 7$^{\rm th}$, 8$^{\rm th}$, 9$^{\rm th}$, 10$^{\rm th}$, and 11$^{\rm th}$ bands.  
We take into account the lattice distortion term of $g_{\rm lp} \equiv 0.1$. 
The Fermi surfaces are also illustrated in the figures. 
We assume $(\Delta^{\rm{dPI}}, \Delta^{\rm{orb}}) = (0.08, 0)$ in (a-e) and 
$(\Delta^{\rm{dPI}}, \Delta^{\rm{orb}}) = (0, 0.08)$ in (f-j). 
}
 \end{center}
\end{figure}

Owing to the lattice distortion term, we obtain 24 eigenstates for each momentum. They are 
described by 12 bands with spin degeneracy when the EO order is absent. The $7^{\rm th} - 11^{\rm th}$ bands cross the Fermi level. 
Thus, we calculate the average spin splitting energy on these Fermi surfaces in the EO state. 
The results are summarized in Table~II. 
It is shown that the $8^{\rm th}$, $9^{\rm th}$, and $10^{\rm th}$ bands show a pronounced spin splitting 
in the dPI-dominated  EO state, while the orbital order gives rise to a large spin splitting 
in the $10^{\rm th}$ and $11^{\rm th}$ bands. 
We understand the difference between the dPI-dominated and orbital-order-dominated  EO states 
on the basis of the orbital character of the bands. The $10^{\rm th}$ and $11^{\rm th}$ bands have the ($d_{yz}$, $d_{zx}$)-orbital characteristic 
and are significantly affected by the orbital order. 
The $7^{\rm th}$, $8^{\rm th}$, $9^{\rm th}$, and $10^{\rm th}$ Fermi surfaces are identified as the $\alpha_1$-, $\alpha_2$-, $\gamma_1$-, 
and $\beta$- Fermi surfaces determined by the ARPES measurements~\cite{Tamai}, respectively, 
while the $11^{\rm th}$ band has not been observed. 

The g-vectors are illustrated in Fig.~9. We see that the spin splitting vanishes at $|k_x| = |k_y|$ 
when the  EO order is induced by the dPI. On the other hand, the spin splitting is 
enhanced at $|k_x| = |k_y|$ if the orbital order is the main cause of the  EO order. 
In both cases, the symmetry of the spin texture is $k_y \, \hat{x} + k_x \, \hat{y}$. 
Thus, our observations for $g_{\rm lp}=0$ (Sects.~2 and 3) are not qualitatively altered by the lattice distortion.

\section{Signature of Electric Octupole Order in the Magnetic Field}

Although we have not considered the effect of the magnetic field, the electronic nematic order in Sr$_3$Ru$_2$O$_7$ occurs  
in the magnetic field of approximately $6 - 8$ Tesla~\cite{Sr3Ru2O7_review}. 
Because the magnetic field lifts the spin degeneracy through the Zeeman effect, 
it may smear out the spin splitting due to the  EO order. 
However, we would examine another signature of the  EO order in the magnetic field, 
which is the asymmetric band structure caused by the cooperation of the  EO order and magnetic field.

\begin{figure}[htb]
 \begin{center}
\hspace*{-2mm}
   \includegraphics[width=7.5cm]{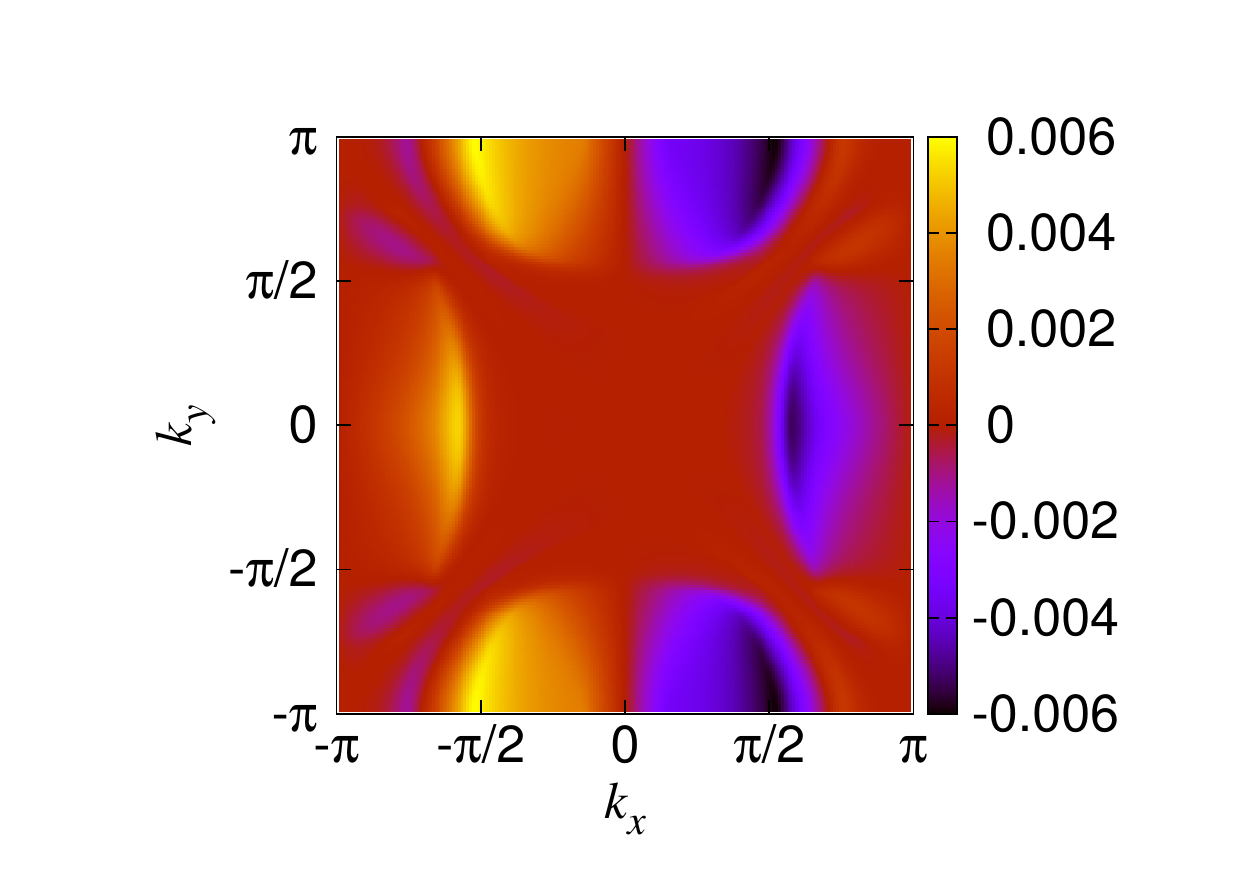}
  \caption{(Color online) 
Asymmetry in the band structure of the $4^{\rm th}$ band, $E_8(k_x, k_y) - E_8(-k_x, k_y)$, in the magnetic field along the [010]-axis. 
We adopt $g_{\rm lp}=0$ and assume $(\Delta^{\rm{dPI}}, \Delta^{\rm{orb}}) = (0.004, 0)$ and $\hh = 0.002 \, \hat{y}$ 
so as to be consistent with the nematic order transition temperature $T_{\rm nematic} \sim 1$ K 
and with the magnetic field $H \sim 6$ T in Sr$_3$Ru$_2$O$_7$. 
}
 \end{center}
\end{figure}

The effective multi-band model [Eq.~(\ref{band-basis})] is useful for discussing the asymmetry in the band structure. 
The Zeeman coupling term is taken into account by replacing the band-dependent g-vector $\g_{j}({\bm k})$ with 
$ \g_{j}({\bm k}) - \frac{1}{2} g^{\rm J}_j \, \mu_{\rm B} \HH$, where $g^{\rm J}_j$ is the Lande g-factor of the $j$-th band 
and $\HH$ is the magnetic field.  
Then, the single-particle energy is obtained as 
\begin{eqnarray}
\hspace{-8mm}
E_{2j/2j-1}(\k) &=& \xi_{j}({\bm k}) \pm \left| \g_{j}({\bm k}) - \frac{1}{2} g^{\rm J}_j \, \mu_{\rm B} \HH \right|
\\\hspace{-8mm}
&\simeq& 
\xi_{j}({\bm k}) \pm \left| \g_{j}({\bm k}) \right| \mp \frac{1}{2} g^{\rm J}_j \, \mu_{\rm B} \left(\hat{\g}_{j}(\k) \cdot \HH \right),  
\end{eqnarray}
where $\hat{\g}_{j}(\k) = \g_{j}(\k)/|\g_{j}(\k)| = \SS_{2j}^{\rm av}(\k)/|\SS_{2j}^{\rm av}(\k)|$.
The last term in Eq.~(17) gives rise to the asymmetry in the band structure 
as a consequence of the antisymmetric g-vector, $\g_{j}(\k) = - \g_{j}(-\k)$. 
Because of the symmetry of the g-vectors 
$\g_j({\bm k}) \, \simeq \, k_y \, \hat{x} \, + \, k_x \, \hat{y} \, + \, \gamma \, k_x \, k_y \, k_z \, \hat{z} $,  
we obtain the asymmetry, 
\begin{eqnarray}
&&\hspace{-9mm}
E_i(k_x, k_y, k_z) \ne E_i(k_x, k_y, -k_z) = E_i(k_x, -k_y, k_z) = E_i(-k_x, k_y, k_z)
\nonumber \\ && \hspace{45mm}
%\hspace{5mm} {\rm for} \hspace{1mm} \HH \parallel [001], 
\end{eqnarray}
for $\HH \parallel [001]$, 
\begin{eqnarray}
&&\hspace{-9mm}
E_i(k_x, k_y, k_z) = E_i(k_x, k_y, -k_z) = E_i(-k_x, k_y, k_z) \ne E_i(k_x, -k_y, k_z)
\nonumber \\ && \hspace{45mm}
%\hspace{5mm} {\rm for} \hspace{1mm} \HH \parallel [100], 
\end{eqnarray}
for $\HH \parallel [100]$, and 
\begin{eqnarray}
&&\hspace{-9mm}
E_i(k_x, k_y, k_z) = E_i(k_x, k_y, -k_z) = E_i(k_x, -k_y, k_z) \ne E_i(-k_x, k_y, k_z)
\nonumber \\ && \hspace{45mm}
%\hspace{5mm} {\rm for} \hspace{1mm} \HH \parallel [010], 
\end{eqnarray}
for $\HH \parallel [010]$. 
According to Eq.~(16), the field angle dependence of the asymmetric band structure would clarify the 
spin texture generated by the EO order. This is a feasible experimental test for the EO order.

The above illustration based on the effective multi-band model is indeed demonstrated by adding the Zeeman term to the 
total Hamiltonian, as $H = H_{\rm{0}} + H_{\rm{dPI}} + H_{\rm{orb}} + H_{\rm Zeeman}$ with $H_{\rm Zeeman} = - \sum_{i} \hh \cdot (2 \SS_i + \LL_i)$. 
Figure~10 shows the asymmetry [Eq.~(20)] in the $4^{\rm th}$ band caused by the magnetic field along the [010]-axis. 
Equations~(18) and (19) have also been verified.

The asymmetry in the band structure is regarded as the emergence of the $p$-wave charge nematic order, whose order parameters are defined as 
\begin{eqnarray}
\label{charge_nematic_order}
&& \hspace*{-8mm}  
\Delta_{\rm pcn}^{x,y,z} = \sum_{l=A,B} \sum_{m=1}^{3} \sum_{\k, s, s'} \sin k_{x,y,z} 
\left< c_{\bm{k}msl}^{\dagger} \, c_{\bm{k}ms'l}  \right>. 
\end{eqnarray}
Generally speaking, the $p$-wave charge nematic order parameter is finite when the space inversion symmetry, 
time-reversal symmetry, and spin SU(2) symmetry are broken.

\section{Summary and Discussion}

We have studied the EO order, which is an odd-parity high-rank multipole order, in the itinerant electron system. 
We showed that the antiferro stacking of the local quadrupole moment in bilayer systems is regarded as an EO order 
from the viewpoint of symmetry.  
Interestingly, the $p$-wave spin nematic order is induced by the spin-orbit coupling. 
Considering the electronic nematic state in the bilayer ruthenate Sr$_3$Ru$_2$O$_7$ as a typical example, 
we elucidated the signatures of the EO order. 

It has been shown that the spin splitting appears in the band structure as a consequence of the spontaneous inversion 
symmetry breaking in the  EO state. 
The spin texture symmetry in Sr$_3$Ru$_2$O$_7$ has been clarified on the basis of the point group symmetry, 
and the spin texture has been calculated by assuming the antiferro stacking of the dPI and orbital order. 
We also showed that the spin texture would be experimentally identified by investigating the asymmetric band structure 
in the magnetic field and its field angle dependence. 

It is expected that intriguing transport phenomena and magnetoelectric effects occur 
as in the noncentrosymmetric metals~\cite{NCSC_chap8}. 
As for Sr$_3$Ru$_2$O$_7$, the enhanced anisotropy of resistivity in a tilted magnetic field~\cite{Sr3Ru2O7_nematic_2, Sr3Ru2O7_review} 
may be attributed to the asymmetric band structure in the EO state. The large residual resistivity in the 
electronic nematic state~\cite{Sr3Ru2O7_nematic_2, Sr3Ru2O7_review} may occur through the domain formation of EO or EQ order. 
We will study the charge and spin transport in the EO state in the future.

Finally, we would like to stress that the strongly correlated electron systems on the locally noncentrosymmetric crystals 
are the platform for realizing the odd-parity multipole order, which is barely stabilized in the even locally centrosymmetric crystals. 
Generally speaking, the antiferro alignment of the even-parity multipole in the unit cell is regarded as an 
odd-parity multipole order. For instance, the ``antiferromagnetic'' order in the unit cell induces the magnetic quadrupole moment 
accompanied by the $p$-wave charge nematic order.~\cite{Magnetic_Quadrupole} 
Such magnetic structure is indeed realized in the zigzag chain structure of CeRu$_2$Al$_{10}$ and related materials.~\cite{CeRu2Al10} 
We also look at the staggered magnetic quadrupole order in the multilayer high-$T_{\rm c}$ cuprate superconductors, 
where the antiferromagnetic moment with $\Q = (\pi,\pi)$ changes the sign between the layers.~\cite{Cuprates}
As we discussed in this paper, the ``antiferro'' order of the local electric 
quadrupole in the unit cell induces the EO order, 
and it may be realized in Sr$_3$Ru$_2$O$_7$.~\cite{Sr3Ru2O7_review} 
As another example, the ``antiferro'' quadrupole order has been observed in PrIr$_2$Zn$_{20}$ for which 
the heavy-fermion superconductivity and non-Fermi liquid behaviors are attracting interest~\cite{Onimaru}. 
The symmetry of the odd-parity multipole and induced nematic order in PrIr$_2$Zn$_{20}$, and its intriguing properties 
will be discussed elsewhere. 
Higher-rank odd-parity multipole orders can also be formed similarly, and an odd-parity nematic order 
is induced by the spin-orbit coupling. 
Other quantum phases with broken inversion symmetry have also been investigated in a recent study~\cite{Hayami_2} 
on the basis of this idea. 
This mechanism would apply to vast materials.

\section*{Acknowledgements}
The authors are grateful to N. Arakawa and R. Shiina for fruitful discussions. 
This work was supported by 
a Grant-in-Aid for Scientific Research on Innovative Areas ``Topological Quantum Phenomena'' 
(No. 25103711) from MEXT Japan, and by a Grant-in-Aid for Young Scientists (No. 24740230) from JSPS. 
Part of the numerical computation in this work was carried out 
at the Yukawa Institute Computer Facility.

%%%%%%%%Bibiography Style File for JPSJ %%%%%%%%%%%%%
% Released on November 15, 1996: Version 1.00       %
% Copyright (C) 1996 by Shinsaku Fujita,            %
%                             all rights reserved.  %
%%%%%%%%%%Bibliography%%%%%%%%%%%%%%%%%%%%%%%%%%%%%%%

% Produces the bibliography via BibTeX.

%\begin{thebibliography}{9}
%\bibitem{jpsj} The abbreviation for JPSJ must be ``J. Phys. Soc. Jpn." in the reference list.
%\bibitem{instructions} More abbreviations of journal titles are listed in ``Instructions for Preparation of Manuscript".
%\end{thebibliography}


\begin{thebibliography}{10}
\bibitem{Comment1}
The antisymmetric spin-orbit coupling is often adopted to describe the 
spin-split band in the noncentrosymmetric metal~\cite{NCSC}. 
That is also an effective spin-orbit coupling, which originates from the 
LS-coupling and odd-parity hybridization terms~\cite{a,Nakamura_STO,Yanase_NCSC}.
\end{thebibliography}
\end{document}